\documentclass{aa}
\usepackage{amssymb}
\usepackage{amsmath}
\usepackage{xspace}
\usepackage{graphicx}
\usepackage{enumerate}   
\usepackage{color,xcolor}
\usepackage{txfonts}

\usepackage{natbib}
\usepackage[colorlinks=true,citecolor=blue,urlcolor=blue,linkcolor=blue]{hyperref}

\def\gr{$\gamma$-ray}
\def\GR{$\gamma$ ray}

\begin{document}

\title{New limit on high Galactic latitude PeV \gr\ flux from Tibet AS$\gamma$ data 
}

\author{A.Neronov$^{1,2}$, D.Semikoz$^{1,3}$ and Ie.Vovk$^{4}$}
\institute{
    Université de Paris, CNRS, Astroparticule et Cosmologie, F-75013 Paris, France
    \and Astronomy Department, University of Geneva, Ch. d'Ecogia 16, 1290, Versoix, Switzerland
    \and Institute for Nuclear Research of the Russian Academy of Sciences, 60th October Anniversary st. 7a, 117312, Moscow, Russia
    \and Institute for Cosmic Ray Research, The University of Tokyo, 5-1-5 Kashiwa-no-Ha, Kashiwa City, Chiba, 277-8582, Japan
}

\authorrunning{A.Neronov et al}
\titlerunning{High Galactic Latitude flux limit}
\abstract
{
    The Tibet AS$\gamma$ collaboration has recently reported the detection of \gr s with energies up to Peta-electronvolt from parts of the Galactic plane. We note that the analysis of $\gamma$-ray  
    flux  by the Tibet-AS$\gamma$ experiment also implies an upper bound on the diffuse $\gamma$-ray flux from high Galactic latitudes $(|b|>20^\circ)$ in the energy range between 100 TeV and 1~PeV. This bound is up to an order of magnitude stronger than previously derived bounds from GRAPES3, KASCADE, and CASA-MIA experiments. We discuss the new Tibet-AS$\gamma$ limit on the high Galactic latitude \gr\ flux in the context of possible mechanisms of multi-messenger (\gr\ and neutrino) emission from nearby cosmic ray sources, dark matter decays, and the large-scale cosmic ray halo of the Milky Way. 
}
\maketitle

\section{Introduction}
Diffuse GeV-TeV \gr\ flux from the sky  is dominated by emission from interactions of cosmic rays in the interstellar medium \citep{fermi_diffuse12,fermi_models,Lipari2018,neronov_diffuse}. In the energy range below several hundred Giga-electronvolts, high Galactic latitude diffuse flux  also has a sizeable contribution from extragalactic sources, but this contribution vanishes at higher energies because of the effect of absorption of extragalactic \gr s during their propagation through the intergalactic medium \citep{gould,Franceschini:2008tp}. 

High Galactic latitude flux has been measured up to several Tera-electronvolts in energy by the Fermi/LAT telescope \citep{neronov_diffuse}. It is dominated by emission from the local interstellar medium, but can also contain contributions from  nearby very extended sources \citep{Neronov:2018ibl,Bouyahiaoui:2018lew}. The diffuse emission from the local interstellar medium has two main contributions: from interactions of cosmic ray protons and atomic nuclei and from inverse Compton emission by cosmic ray electrons. The modelling of these two components suggests that pion decay flux from interactions of protons and nuclei dominates over the inverse Compton flux in this part of the sky \citep{Lipari2018}. The inverse Compton flux above several Tera-electronvolts is further suppressed by the decrease in the interaction cross section in the Klein-Nishina regime and by the softening of the cosmic ray electron spectrum above Tera-electronvolt \citep{2008PhRvL.101z1104A,PhysRevLett.120.261102,DAMPE:2017fbg}.

At even higher energies, diffuse \gr\ flux has been measured from parts of the sky by extensive air shower (EAS) arrays MILAGO, HAWC, and ARGO-YBJ \citep{Abdo:2008if,ARGO-YBJ:2015cpa,Abeysekara_2017}. In this energy range, emission from the Galactic plane generally follows the powerlaw extrapolation of the lower energy flux \citep{neronov_diffuse}, as expected from the flux produced by the cosmic rays with a powerlaw spectrum. No measurements of the \gr\ flux at high Galactic latitudes have been reported so far.  In the absence of detection, the EAS experiments have previously reported upper bounds on the all-sky \gr\ flux, assuming its near isotropy. The tightest reported bounds in the 1 PeV range are from the KASCADE \citep{Apel:2017ocm} and CASA-MIA \citep{Chantell:1997gs} experiments. At somewhat lower energies in the 0.01-0.1~PeV range, the best bounds come from the GRAPES3 EAS experiment~\citep{grapes3} and the HEGRA Cherenkov telescope array \citep{Aharonian:2001ft}. The \gr\ flux at 100 TeV
can be produced by cosmic ray protons and nuclei of energies far above 100 TeV in the energy range of the `knee' of the cosmic ray spectrum. The measurement of the \gr\ flux in this energy range can thus yield information on the nature of the knee and on the location and properties of  nearby cosmic accelerators
boosting particle energies beyond 1 PeV, the so-called pevatrons. If the particles responsible for the \gr\ emission are protons and atomic nuclei, the 1 PeV \gr\ emission from the pevatrons is accompanied by the emission of neutrinos, with comparable luminosity and spectral characteristics. The detection of the 0.1-1~PeV diffuse \gr\ flux can thus be used to constrain the Galactic component of the astrophysical neutrino flux \citep{icecube} of uncertain origin.

The Tibet-AS$\gamma$ experiment has recently reported detection (rather than an upper bound) of diffuse emission from the Galactic plane in the energy range up to 1~PeV \citep{Amenomori2021}. 
The better sensitivity of Tibet-AS$\gamma$ and higher energy reach, compared to MILAGRO, HAWC, and ARGO-YBJ, can be explained by its more efficient suppression of background EAS produced by protons and atomic nuclei.  The supplementary material for the Tibet-AS$\gamma$ publication shows the efficiency of  the suppression of the charged cosmic ray background reaching the $\sim 10^{-6}$ level of the all-particle cosmic ray flux at 1~PeV. The Tibet-AS$\gamma$ detection of the Galactic plane can be used to constrain the population of cosmic ray electrons \citep{2021arXiv210409491F} and cosmic ray protons and nuclei  \citep{Koldobskiy:2021cxt} in the Galactic disk.

In what follows we note that the Tibet-AS$\gamma$ analysis has another important implication: it constrains the \gr\ flux from the sky outside the Galactic plane. We extracted{ A\&A uses the past tense to describe specific methods used in a paper, and the present tense to describe general methods and the findings of recent papers. See Sect. 6 of the language guide https://www.aanda.org/for-authors/language-editing/6-verb-tenses.
Please review my edits to ensure this was carried out appropriately throughout your paper.} an upper limit on the high Galactic latitude flux from the Tibet-AS$\gamma$ measurements and explore its implications for models of high-energy particle acceleration in the Milky Way and for the models of super-heavy dark matter. 

\section{Tibet-AS$\gamma$ limit on the high Galactic latitude \gr\ flux}

\begin{figure}
\includegraphics[width=\linewidth]{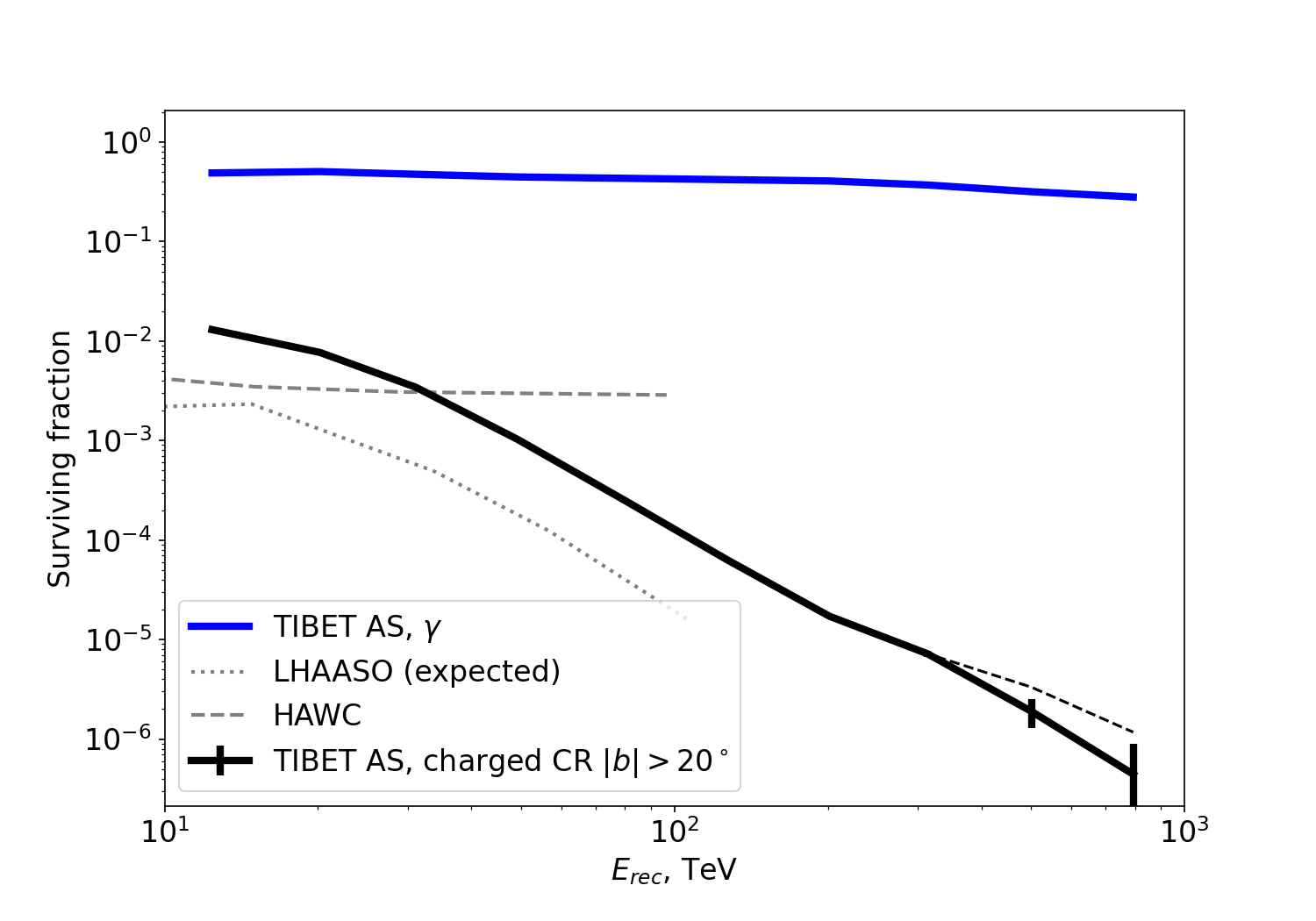}
\caption{Surviving fraction of proton and nuclei EAS in Tibet~AS$\gamma$ experiment after \gr-like event selection \citep{Amenomori2021}. The dashed black line shows the upper limit on the total number of surviving \gr- like events in the $|b|>20^\circ$ region. The blue line shows the efficiency of the selection of the \gr\ EAS. For comparison, grey dashed and dotted lines show the efficiencies of the suppression of the hadronic EAS background in HAWC~\citep{2017ApJ...843...39A} and LHAASO~\citep{lhaaso}. }
\label{fig:fraction}
\end{figure}

The Tibet-AS$\gamma$ array achieves a high efficiency of rejection of the cosmic ray proton and nuclei background due to a large size muon detector. Hadronic EASs are typically muon-rich, whereas \GR-\ and electron-induced showers are muon-poor. Discriminating between a \gr ~and  an electron as well as a proton and a nuclei EAS becomes progressively more efficient with increasing energy because of larger muon statistics. This is shown in Fig. \ref{fig:fraction} which compares the efficiency of hadronic EAS background rejection in Tibet~AS$\gamma$ to that of HAWC \citep{2017ApJ...843...39A}  and LHAASO \citep{lhaaso}. The blue curve in Fig. \ref{fig:fraction} shows the efficiency of the selection of the \gr\ EASs for the same cut selection that suppresses the hadronic EAS background down to the black curve level~\citep{Amenomori2021}.

Fig. \ref{fig:fraction} shows that  starting from 30 TeV, the background suppression of Tibet~AS$\gamma$ experiments starts to be better than that of HAWC. This is explained by the absence of a muon detector in HAWC. Similarly to Tibet~AS$\gamma$, the LHAASO experiment \citep{Bai:2019khm}, which is now under way, will use muon detectors for gamma-hadron separation. This explains a similar energy dependence of LHAASO and Tibet~AS$\gamma$ background suppression efficiencies.  

The non-negligible fraction of the EAS that are not rejected by the cuts on `gamma-like' events can, in principle, be real \gr\ events. It is, however, not possible anymore to distinguish the remaining hadronic EAS from the real \gr\ EAS based on the EAS parameters. The presence of real \gr\ events from a specific astronomical source can be spotted by comparing the statistics of events coming from the source direction with that for events from other directions on the sky. This has been done in the analysis of \cite{Amenomori2021}, leading to the detection of the signal from the Galactic plane. 

Comparing of the `on-source' and `off-source' event statistics is, however, not possible for the source spanning a large part of the sky because there is no `off-source' region in this case. It might still be possible to detect sources in this regime if the observed event statistic shows an excess compared to the expectations based on the known hadronic cosmic ray spectrum and the efficiency of rejection of the hadronic cosmic rays derived from Monte-Carlo modelling of detector performance. In the absence of such modelling, only upper limits on the \gr\ flux from the large sky region can be derived, assuming that all the events remaining after the cuts are possible \GR s. 

\begin{figure}
\includegraphics[width=\linewidth]{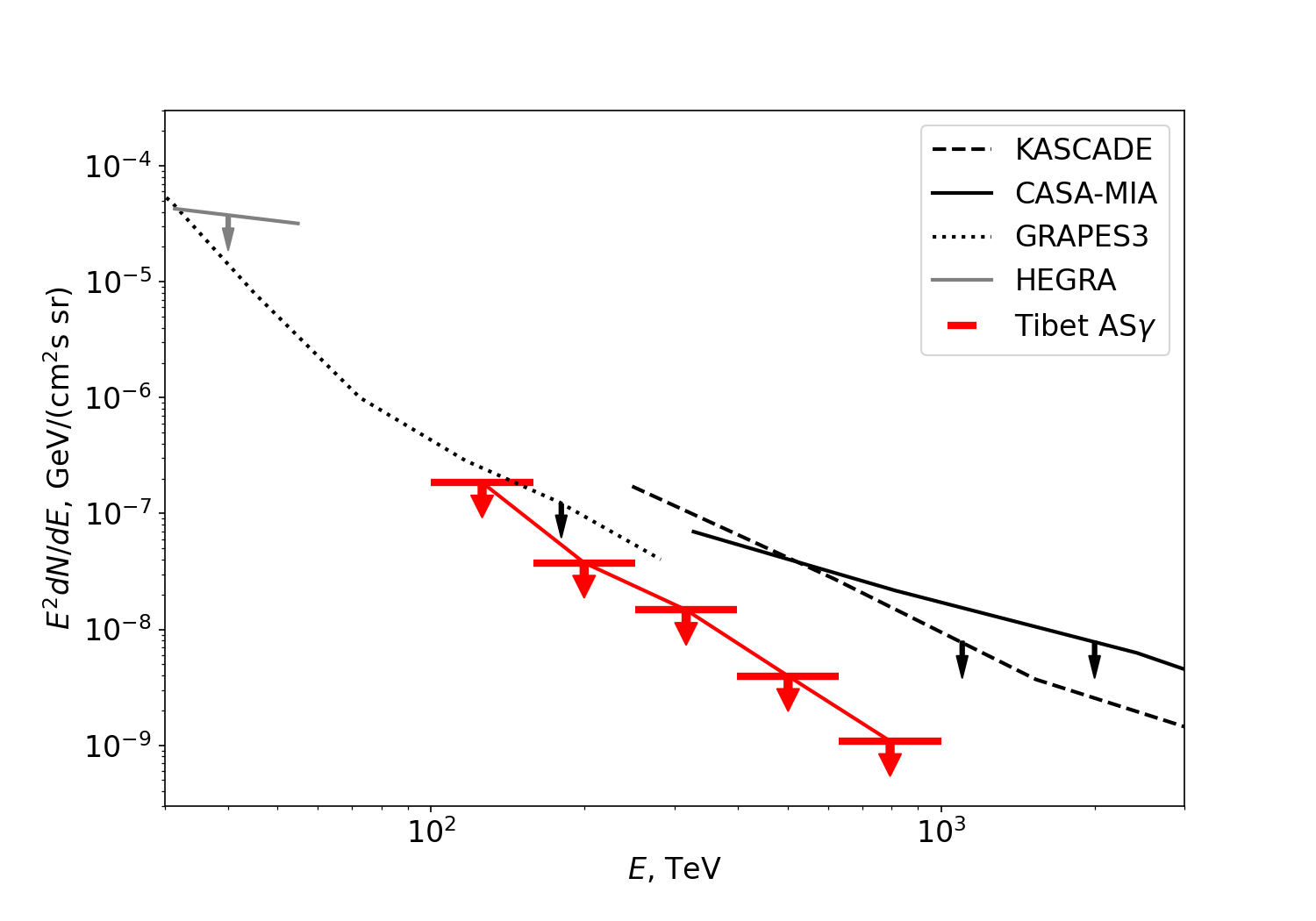}
    \caption{Tibet AS$\gamma$ limit on high Galactic latitude \gr\ flux derived from~\citet{Amenomori2021}, compared to previously reported limits from KASCADE \citep{Apel:2017ocm}, CASA-MIA \citep{Chantell:1997gs}, GRAPES3 \citep{grapes3}, and HEGRA \citep{Aharonian:2001ft} experiments. }
    \label{fig:ul}
\end{figure}

This approach gives the upper limit on the \gr\ flux in the sky region $|b|>20^\circ$ shown in Fig. \ref{fig:ul}. To derive this upper limit, we used the measurement of the all-particle cosmic ray flux by the Tibet~AS$\gamma$ experiment \citep{tibet_cr} multiplied by the surviving event fraction of hadronic EASs and divided by the efficiency of the selection of \gr\ EASs.
The last two bins have been corrected for the effect of low statistics on the surviving signal. The derived limits thus correspond to the same 719-day live time as in~\citet{Amenomori2021}.

Fig. \ref{fig:ul} provides a comparison of the upper limit on the diffuse flux from the sky region $|b|>20^\circ$ with previously reported regions on the sky-averaged \gr\ flux. One can see that in the 0.3-1~PeV energy range, the limit derived from Tibet~AS$\gamma$ data is an order of magnitude better than previous limits from KASCADE~\citep{Apel:2017ocm} and CASA-MIA~\citep{Chantell:1997gs} experiments. At lower energies, the Tibet~AS$\gamma$ limit is comparable to that from the GRAPES3 experiment~\citep{grapes3}. 

\section{Implications of the revised bound}

\begin{figure}
    \includegraphics[width=\linewidth]{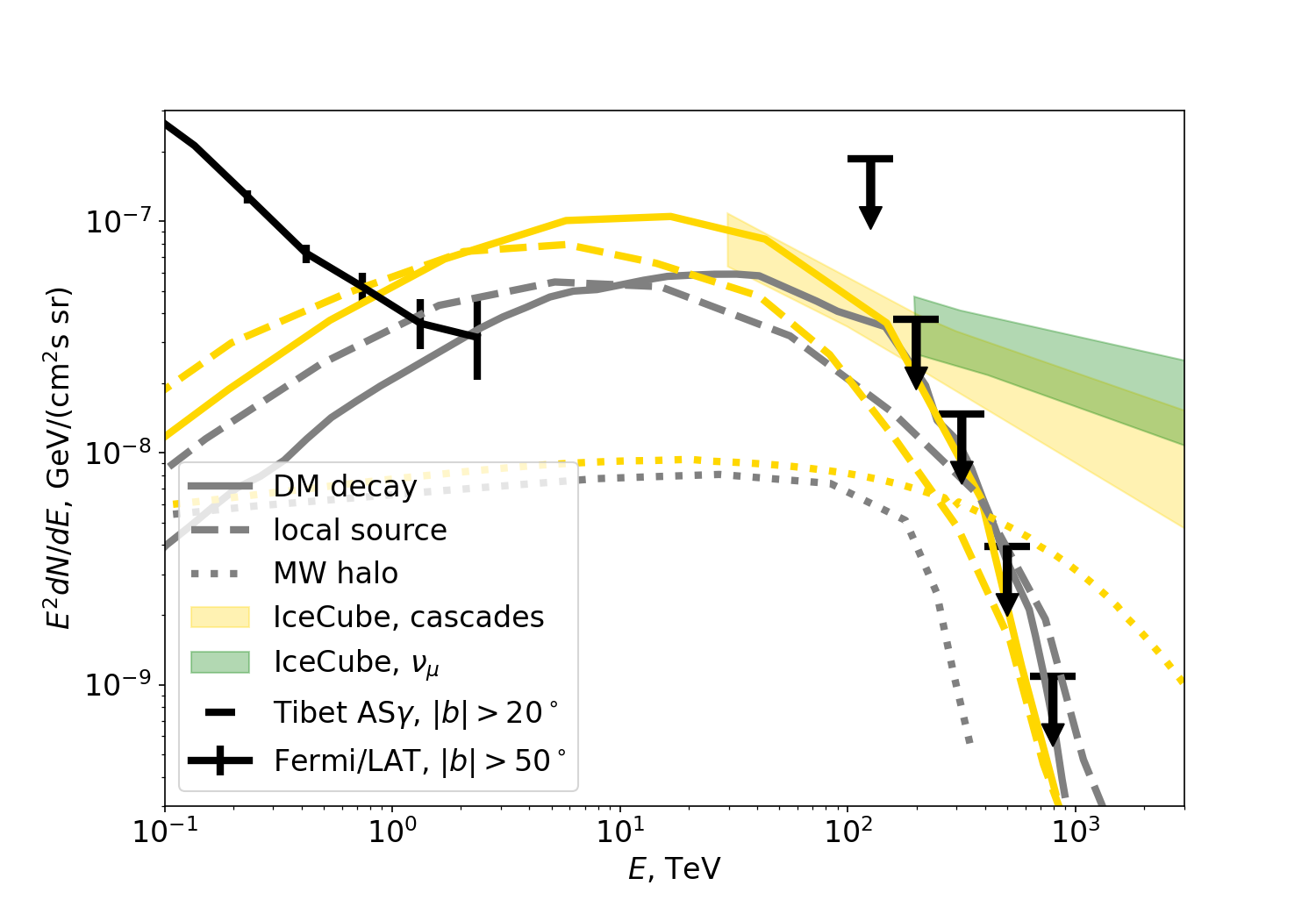}
    \caption{
        Tibet AS$\gamma$ limit on high Galactic latitude \gr\ flux compared to Fermi/LAT measurements at lower energies from \cite{neronov_diffuse} and to model predictions of neutrino (yellow lines) and \gr\ (grey lines) fluxes  for the local cosmic ray source \citep{2020PhRvD.101l3023B} (dashed lines), dark matter decay \citep{2018PhRvD..98b3004N} (solid), and 100 kpc scale halos of the Milky Way \citep{2021arXiv210105016R} (dotted). Yellow and green areas show the astrophysical neutrino spectrum measurement in cascade  \citep{Aartsen:2020aqd} and muon \citep{2019ICRC...36.1017S} channels.
    }
    \label{fig:models}
\end{figure}

An improvement to the upper limit on the diffuse \gr\ flux from the high Galactic latitude region, provided by the Tibet~AS$\gamma$ experiment, can be used to probe a range of models of \gr\ and neutrino emission from particle acceleration processes and dark matter in the Milky Way. Fig. \ref{fig:models} shows a comparison of the new upper limit in the 0.1-1~PeV range with measurement of the high Galactic latitude flux in the region $|b|>50^\circ$ with Fermi/LAT at the energies below 3~TeV \citep{neronov_diffuse} and with IceCube measurements of the astrophysical neutrino flux in muon \citep{2019ICRC...36.1017S} and cascade  \citep{Aartsen:2020aqd}  channels. 

The astrophysical muon neutrino flux measured above 200 TeV 
has a harder powerlaw spectrum which is shown by the green shaded range ~\citep{Aartsen:2016xlq,Stettner:2019tok}. Its overall energy flux is lower than the isotropic diffuse \gr\ flux measured by Fermi-LAT below 1 TeV~\citep{TheFermi-LAT:2015ykq}. This is consistent with  a possibility that both the Peta-electronvolt energy scale neutrinos and Giga-electronvolt range \gr s are produced by the same extragalactic source population~\citep{Murase:2013rfa}. Blazars are possible candidates for \gr\ and neutrino sources and they provide a major contribution to the  high-energy part of Fermi LAT diffuse \gr\ background~\citep{Abdo:2010nz,Neronov:2011kg,DiMauro:2013zfa,TheFermi-LAT:2015ykq}. However, \gr\ bright blazars do not offer a major contribution  of IceCube muon neutrino flux \citep{Aartsen:2016lir,Neronov:2016ksj}. Gamma-ray loud blasars still can provide up to 20 \% of total muon neutrino flux according to \cite{Giommi:2020hbx}. A piece of  evidence for a significant  contribution from radio-brightest blazars has been reported recently~\citep{Plavin:2020emb,Plavin:2020mkf}. Otherwise, non-blazar active galactic nuclei (AGN) \citep{Smith:2020oac} and ultra-high energy cosmic ray sources~\citep{Kachelriess:2017tvs} can contribute to the neutrino flux. Uncertainties regarding the modelling of the extragalactic neutrino source populations still leave a possibility that part of the neutrino flux comes from the local Galaxy. The new Tibet-AS$\gamma$ bound derived above severely limits a possible Galactic component of the neutrino flux in the Peta-electronvolt energy range. Fig. \ref{fig:models} shows that no more than 10\% of the neutrino flux above 1 PeV detected from the high Galactic latitude region can come from the region within 10~kpc around the Solar System. The \gr\ flux from larger distances is attenuated by the effect of pair production on the cosmic microwave background. The mean free path of PeV \gr s with respect to the pair production is $\lambda_{\gamma\gamma}\simeq 10$~kpc \citep{gould}. If as of yet uncertain source(s) of high Galactic latitude neutrino flux is(are) situated at the distances $D\gtrsim 2\lambda_{\gamma\gamma}$ for which $exp(-D/d_{\gamma\gamma})\lesssim 0.1$, the \gr\ flux accompanying the neutrino flux can still be suppressed down to the level of the Tibet-AS$\gamma$ bound. 

The astrophysical neutrino flux has also been detected at lower energies in the 'cascade' channel, down to $\sim 10$~TeV, where it appears to have a softer spectrum~\citep{Aartsen:2020aqd,Abbasi:2020jmh} shown by the yellow range in Fig. \ref{fig:models}. Its spectral properties suggest that it is produced in proton-proton or nuclei-nuclei interactions, rather than $p-\gamma$ interactions that are typically characterised by an energy threshold in the range above 100~TeV and the hard neutrino spectrum below the threshold energy. In this case the soft neutrino spectrum is expected to extend down to the 1 GeV range. It has been suggested that the softer spectrum neutrinos have a Galactic origin \citep{Neronov:2013lza,Neronov:2014uma,Neronov:2015osa}. However, a large signal from the Galactic plane is inconsistent with arrival directions of  neutrino events \citep{ANTARES:2018nyb}, suggesting that the latter come from either hidden extragalactic sources \citep{Murase:2015xka} or from  galactic sources distributed away from the Galactic plane \citep{Taylor:2014hya,Kachelriess:2018rty}. In this last case one should see the unavoidable diffuse gamma-ray background at high galactic latitudes.
Fig. \ref{fig:models} shows that Fermi/LAT high latitude flux and IceCube astrophysical neutrino energy fluxes are in fact comparable (albeit sampled in different energy ranges). This suggests that a common process producing neutrinos and \GR s may indeed provide non-negligible contribution to both  \gr\ and neutrino diffuse fluxes. The new Tibet-AS$\gamma$ limit at 100~TeV still does not rule out the possibility of the existence of a \gr\ flux at the level of neutrino flux up to the 100~TeV range. 

Several models for the common process generating a high Galactic latitude  neutrino and \gr\ fluxes have been proposed, including the possibility of  local cosmic ray sources~\citep{Kachelriess:2015oua,Kachelriess:2017yzq,2018PhRvD..98b3004N,Bouyahiaoui:2018lew,Bouyahiaoui:2020rkf}, the decay of super-heavy dark matter particles~\citep{Berezinsky:1997hy,Feldstein:2013kka,Esmaili:2013gha,2018PhRvD..98b3004N,Kachelriess:2018rty}, as well as a large-scale cosmic ray halo of the Milky Way galaxy~\citep{Taylor:2014hya,Kalashev:2016euk,Blasi:2019obb,2021arXiv210105016R}. These models have been previously constrained by the Fermi/LAT and IceCube data, but not by the \gr\ data in the IceCube energy range. The new Tibet~AS$\gamma$ upper bound on the high Galactic latitude flux starts to constrain these models. In particular, predictions of the model of \GR s from dark matter particle decays by \citet{2018PhRvD..98b3004N} exceed the Tibet~AS$\gamma$ limit. The model can be made consistent with Tibet~AS$\gamma$ data only if the mass of the dark matter particle is lower than $M_{DM}c^2<3$~PeV. The local source model is not yet constrained, but its prediction is right at the Tibet~AS$\gamma$ upper bound on the \gr\ flux. 

Even though the large-scale halo model has been initially proposed as an explanation for the astrophysical neutrino signal~\citep{Taylor:2014hya}, its flux has been strongly constrained by the Fermi/LAT measurement of the isotropic \gr\ background, so that the neutrino flux from the halo is far below the measured astrophysical neutrino flux~\citep{2021arXiv210105016R}. As such, the model is also consistent with the Tibet~AS$\gamma$ data. 

Measurements of the diffuse \gr\ flux at a high Galactic latitude will be crucially improved by LHAASO~\citep{lhaaso}. The potential of LHAASO is clear from Fig. \ref{fig:fraction} which  shows that LHAASO's efficiency in suppressing the hadronic EAS background will be 6~times lower than that of Tibet~AS$\gamma$. This should a substantial improvement to the upper bound or detection of the signal in 0.3-1~PeV range where the Tibet~AS$\gamma$ bound is close to the model predictions and where the maximal possible flux level is close to the level of the neutrino flux. 

\begin{acknowledgements}
Authors are grateful to Masato Takita and Kazumasa Kawata from Tibet AS$\gamma$ collaboration for useful discussions during the preparation of this manuscript. Work of A.N. and D.S. was supported in part by the Ministry of science and higher education of Russian Federation under the contract 075-15-2020-778 in the framework of the Large scientific projects program within the national project `Science'.
\end{acknowledgements}

\bibliographystyle{aa}
\bibliography{refs.bib}

\end{document}